\documentclass[12pt]{article}

\usepackage{amsmath,eucal,epsf,amsfonts}

\newcommand{\Dc}{\partial}
\renewcommand{\Re}{\operatorname{Re}}

\newcommand{\residuo}{\operatorname {res}}
\newcommand{\fp}{\operatorname{FP}}
\newcommand{\tr}{\operatorname{tr}}

\newcommand{\D}{\ensuremath{{\not\!\!D}}}
\newcommand{\Sl}{\ensuremath{\underset {y\rightarrow x}{
\operatorname{Sch-lim}}}}

\author{R.E.Gamboa Sarav\'\i $^1$ \and M.A.Muschietti$^2$ \and J.E.Solomin$^2$ \\ \hfill \and
{ Facultad de Ciencias Exactas, UNLP}\\
$^1$Departamento de F\'{\i}sica\\
$^2$Departamento de Matem\'atica}
\title{On gauge invariant regularization of fermion currents.
\thanks{Partially supported by CONICET, Argentina. } }
\date{ }

\begin{document}

\maketitle
\begin{abstract}
We compare  Schwinger and complex powers methods to construct
 regularized fermion currents. We show that  although
 both of them are gauge invariant they not always yield the same
 result.

\end{abstract}
\bigskip
\bigskip
e-mail address: quique@dartagnan.fisica.unlp.edu.ar\newline
%
\eject

A difficulty specific to quantum field theories is the occurrence
of infinities and hence the necessity of regularizing and
renormalizing the theory. Whenever a field theory possesses a
classical symmetry---and hence a conserved current---it is
desirable to have at hand regularization procedures preserving
that symmetry.\footnote{As it is well known, not always it is
possible to preserve all the classical symmetries present
simultaneously  and {\em anomalies} can arise.}

The calculation of vacuum expectation values of vector currents
involves the evaluation of the Green function for the particle
fields at the diagonal, so a regularization is required. In a
classical paper J. Schwinger introduced a point splitting method
to regularize fermion currents maintaining gauge symmetry on the
quantum level \cite{Sch}.

More recently, the so called $\zeta$-function method, based on
complex powers of pseudodifferential operators \cite{See}, has
proved to be a very valuable gauge invariant regularizing tool
(see for example \cite{Haw}). Some time ago, we used it to get
fermion currents in 2 and 3 dimensional models \cite{Jour}.

It is the aim of this work to compare the results obtained by the
above mentioned methods.

\bigskip

Let $\D=i\!\not \! \partial +\!\! \not\! \!A$ be an Euclidean
Dirac operator coupled with a gauge field $A$  defined on a
n-dimensional compact boundaryless manifold $M$. The operator
$\D$ is elliptic and, since  its principal symbol has only real
eigenvalues, it fulfills  the Agmon cone condition \cite{See}.
Thus, the complex powers $ \D^s$ can be constructed following
Seeley~\cite{See}. For $\Re s <0$ we can write
\begin{equation}\label{Ds}
\D^s:=\dfrac{i}{2\pi}\int_{\Gamma} \lambda^s(\D-\lambda)^{-1}
d\lambda,
\end{equation}
where $\Gamma$ is  a contour enclosing the spectrum of \D, and we
define $ \D^s$ for  $\Re s \geq0$ by using  $ \D^{s+1}=\D^s \circ
\D $.

For each $s \in \mathbb{C}$,  $\D^s$ turns out to be a
pseudodifferential operator  of order $s$ and so, if $\Re s<-n$,
 its Schwartz kernel $K_s(x,y)$ is a
continuous function. The evaluation at the diagonal $x=y$ of this
kernel, $K_s(x,x)$, admits a meromorphic extension to the whole
complex s-plane $\mathbb{C}$, with at most simple poles at $s\in
\mathbb{Z}^-$. This extension will be also denoted by $K_s(x,x)$.

Since $K_{-1}(x,y)$ coincides with the Green function for $x\neq
y$, the finite part of $K_s(x,x)$ at $s=-1$ can be used to obtain
gauge invariant regularized fermion currents \cite{Jour}:
\begin{equation}\label{J}
J_\mu(x) =- \tr \gamma_\mu \left(\underset{s=-1}{\fp}
K_{s}(x,x)\right)
\end{equation}

In order to  compare  this regularizing procedure with
Schwinger's one, it is convenient to consider the kernels
$K_s(x,x)$ in the framework developed in \cite{KV}. Since we are
interested in studying the behaviour of these kernels for
$s\rightarrow -1$, we shall carry out our analysis just for  $-1
\leq \Re s <0$.

By considering the finite expansion (see for instance
\cite{Shubin})
\begin{equation}\label{00}
\sigma( \not \!\!D^{s})\ = \sum_{\ell=0}^{N} {c}_{s-\ell}(x,\xi) +
r_N(x,\xi,s),
\end{equation}
with $N=n-1$, of the symbol of the operator ${\not \!\!D}^{s}$,
with ${c}_{s-\ell}(x,\xi)$ positively homogeneous of degree
$s-\ell$ for $\left|\xi\right|\geq 1 $,  we can write, for  $s
\neq -1$ the Schwartz kernel of this operator as
\begin{equation}\label{01}
K_s(x,y) = \sum_{\ell=0}^{N} H_{-n-s+\ell}\ (x,u) + R_N(x,u,s),
\end{equation}
where $H_{-n-s+\ell}(x,u)$ is the Fourier transform in the
variable $\xi$ of  $\widetilde{c}_{s-\ell}(x,\xi)$, the
homogeneous extension of ${c}_{s-\ell}(x,\xi)$, evaluated at
$u=x-y$,  and consequently u-homoge\-neous of degree $-n-s+\ell$
and $R_N(x,u,s)$ is that of $r_N(x,\xi,s)-\sum_{\ell=0}^{N}
(\widetilde{c}_{s-\ell}-{c}_{s-\ell})(x,\xi)$. Note that
$(\widetilde{c}_{s-\ell}-{c}_{s-\ell})(x,\xi)\equiv0$ for
$\left|\xi\right|\geq 1 $.

Now, for $u\neq0$, simple poles can arise at $s=-1$ in
$H_{-n-s+N}$ and in $R_N(x,u,s)$ \cite{KV}. Since $K_s(x,x-u)$ is
holomorphic in the variable $s$ for $u\neq0$, these poles cancel
each other. In fact, they are just due to the singularity of
$\widetilde{c}_{s-N}(x,\xi)$ at $\xi=0$ and then
\begin{equation}\label{03}
\underset{s=-1}{\residuo}R_N(x,u,s) = -\underset{s=-1}{\residuo}
H_{-n-s+N}\ (x,u).
\end{equation}
Thus, for $u\neq0$, we have for $G(x,y)$, the Green function of
$\D$,
\begin{equation}\label{04}
G(x,y)=\lim_{s\rightarrow-1}\ K_s(x,y)=\sum_{\ell=0}^{N}
G_{-n+1+\ell}\ (x,u) + R_G(x,u),
\end{equation}
with $G_{-n+1+\ell}\ (x,u)= \underset {s\rightarrow-1}{\lim}\
H_{-n-s+\ell}\ (x,u)$ for $\ell<N$, $G_{-n+1+N}\ (x,u)= \underset
{s=-1}{\fp}\ H_{-n-s+N}\ (x,u)$ and $R_G(x,u) = \underset
{s=-1}{\fp}\ R_N(x,u,s)$.

Then, taking into account that, for $s\neq -1$, (see, for
instance \cite{KV})
\begin{equation}\label{05}
K_s(x,x)= R_N(x,0,s) ,
\end{equation}
we have
\begin{equation}\label{06}
\underset{s=-1}{\fp}K_s(x,x)= R_G(x,0) ,
\end{equation}

On the other hand, the fermionic currents regularized according
to Schwin\-ger's prescription are given by \cite{Sch}
\begin{equation}\label{JSch}
J_\mu(x)= -\Sl \tr \left( \gamma_\mu G(x,y)\ e^{i\int_x^y A.dz}
\right),
\end{equation}
where
\begin{equation}
\int_x^y A.dz={-\int_0^1 A_\mu(x-tu)\ u_\mu\  dt},
\end{equation}
and \Sl\ (Schwinger limit) is  the usual limit when it exists, it
vanishes for u-homogeneous functions of negative degree and for
logarithmic ones, and it coincides with the mean value at $|u|=1$
for u-homogeneous functions of zero degree. The exponential
factor was introduced by Schwinger \cite{Sch} in order to
maintain gauge invariance.

 From (\ref{J}), (\ref{06}) and (\ref{JSch}) we see that both methods
 yield  the same result for $J_\mu$ if and only if
\begin{equation}\label{chivo}
\Sl \tr \left( \gamma_\mu \sum_{\ell=0}^{N} G_{-n+1+\ell}\ (x,u)\
e^{i\int_x^y A.dz} \right)=0
\end{equation}
since, being $R_G(x,u)$ continuous at $x=y$,
\begin{equation}\begin{split}
\Sl \tr \left( \gamma_\mu R_G(x,u) \ e^{i\int_x^y A.dz}
\right)&=\Sl \tr \left( \gamma_\mu R_G(x,u)  \right)
\\=\lim_{u\rightarrow0}\tr \left( \gamma_\mu R_G(x,u)
\right)&=\tr \left( \gamma_\mu \underset{s=-1}{\fp}K_s(x,x)
\right).
\end{split}\end{equation}

Now, we shall see how this works  in $n=2$,  $3$ and $4$. By
computing the $ G_{-n+1+\ell}\ (x,u)$'s we shall be able to
establish when (\ref{chivo}) holds and so, when both methods yield
the same regularized currents.

In a local coordinate chart
\begin{equation}
\D= \gamma_\mu D_\mu= \gamma_\mu(i\partial_\mu+A_\mu),
\end{equation}where the algebra of the $\gamma$-matrices is
\begin{equation}
\gamma_\mu\gamma_\nu+\gamma_\nu\gamma_\mu=\delta_{\mu\nu}.
\end{equation}
Its symbol, $\sigma(\not \!\!D;x,\xi)$, is
\begin{equation}
\sigma(\not \!\!D;x,\xi)= -\not \!\xi-\not \!\!A(x).
\end{equation}

The symbol of the resolvent, $\sigma((\D-\lambda)^{-1};x,\xi)$,
has an asymptotic expansion $\sum_\ell
\widetilde{C}_{-1-\ell}(x,\xi,\lambda)$, where
$\widetilde{C}_{-1-\ell}(x,\xi,\lambda)$ is homogeneous in $\xi$
and $\lambda$ of degree  $-1-\ell$ \cite{See}. Then
\begin{equation}
( \not \!\!D-\lambda)^{-1}\ \varphi(x)  \sim
\frac{1}{(2\pi)^{n/2}}\int\sum_\ell
\widetilde{C}_{-1-\ell}(x,\xi,\lambda)\  e^{i\xi.x}\
\widehat{\varphi}(\xi)\  d\xi,
\end{equation}
 Applying \  $  \D-\lambda$ to  Equation (\ref{00})
we get  recursive equations for determining the
$\widetilde{C}_{-1-\ell}(x,\xi,\lambda)$'s:
\begin{equation}
\begin{split}
&-( \not \!\xi+\lambda)\ \widetilde{C}_{-1}(x,\xi,\lambda)=1\\
&\not \!\!D_x \widetilde{C}_{-1-\ell}(x,\xi,\lambda)\  -( \not
\!\xi+\lambda)\ \widetilde{C}_{-1-\ell-1}(x,\xi,\lambda)=0.
\end{split}
\end{equation}
Consequently,
\begin{equation}
\widetilde{C}_{-1-\ell}(x,\xi,\lambda)=\  -\frac{( \not
\!\xi-\lambda)}{\xi^2-\lambda^2}\ \left[\  \not \!\!D_x \  \frac{(
\not \!\xi-\lambda)}{\xi^2-\lambda^2}\right]^\ell.
\end{equation}
Now, from equation (\ref{Ds}),
\begin{equation}\begin{split}
H_{-n-s+\ell}(x,u)= \frac{1}{(2\pi)^n}\int
\widetilde{c}_{s-\ell}(x,\xi)\ e^{i\xi.u}\ d\lambda \ d\xi\\ \\=
\frac{i}{(2\pi)^{n+1}}\int\int_\Gamma
\widetilde{C}_{-1-\ell}(x,\xi,\lambda)\ \lambda^s\ e^{i\xi.u}\
d\lambda \ d\xi,
\end{split}\end{equation}
where  the contour $\Gamma$ can be chosen as shown in Figure 1.
Therefore,
\begin{equation}
\begin{split}
&H_{-n-s+\ell}(x,u)\\ \\&= \frac{-i}{(2\pi)^{n+1}}\int\int_\Gamma
\frac{( \not \!\xi-\lambda)}{(\xi^2-\lambda^2)^{\ell+1}}\ \left[\
\not \!\!D_x \ ( \not \!\xi-\lambda)\right]^\ell\ \lambda^s\
e^{i\xi.u}\ d\lambda \ d\xi\\ \\ &=
\frac{-i}{(2\pi)^{n+1}}\int\int_\Gamma
 \frac{(-i \not
\!\partial_u-\lambda)}{(\xi^2-\lambda^2)^{\ell+1}}\ \left[\  \not
\!\!D_x \ (-i \not \!\partial_u-\lambda)\right]^\ell\ \lambda^s\
e^{i\xi.u}\ d\lambda \ d\xi.
\end{split}
\end{equation}

\begin{figure}[t]\caption{The $\Gamma$ curve in the $\lambda$-plane.}

\epsfysize 5cm \centerline{\epsffile{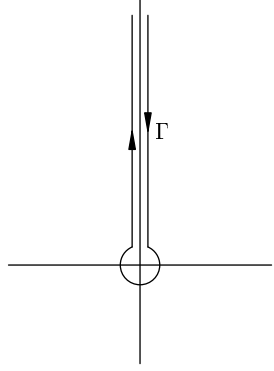}} \end{figure}

Taking into account that, for any polynomial $P(\lambda)$,
\begin{equation}
\begin{split}
&\frac{i}{2\pi}\int_\Gamma \frac{ \lambda^s\ P(\lambda)
}{(\xi^2-\lambda^2)^{\ell+1}}\ \ d\lambda
\\ \\&= \frac{i}{2\pi}\left\{\int_\infty^0
\frac{ (z\ e^{i\frac{\pi}{2}})^s\ P(iz) }{(\xi^2+z^2)^{\ell+1}}\ \
i\ dz\ + \int^\infty_0 \frac{ (z\ e^{-i\frac{3\pi}{2}})^s\ P(iz)
}{(\xi^2+z^2)^{\ell+1}}\ \ i\
dz \right\}\\ \\%
&= \frac{i}{\pi}\ \ e^{-i\frac{\pi}{2}s}\ \sin(\pi s)\
P(-\partial_a) \left[ \int^\infty_0 \frac{ z^s\ \ e^{-iaz}
}{(\xi^2+z^2)^{\ell+1}}\ \ dz \right]_{a=0},
\end{split}
\end{equation}
we can write
\begin{equation}
\begin{split}\label{h1}
H_{-n-s+\ell}(x,u)= \frac{-i}{\pi}\ \ e^{-i\frac{\pi}{2}s}\
\sin(\pi s)( -i\not \!\partial_u+\partial_a)\ \left[\ \not
\!\!D_x \
( -i\not \!\partial_u+\partial_a)\right]^\ell\\
\\ \times \left.\sum^{\ell+1}_{k=0}\frac{(-ia)^k}{k!}\int_0^\infty z^{s+k}
\frac{1}{(2\pi)^n}\int
 \frac{1}{(\xi^2+z^2)^{\ell+1}}\ \ e^{i\xi.u}\ d\xi \ dz
 \right|_{a=0}.
\end{split}
\end{equation}
Now, the integrals in (\ref{h1}) can be performed using the known
identities
\begin{equation}
\begin{split}
\frac{1}{(2\pi)^n}\int (\xi^2+z^2)^s \ e^{i\xi.u}\ d\xi
=\frac{2^{1+s}}{(2\pi)^\frac{n}{2}}\ \frac{1}{\Gamma(-s)}\
\left(\frac{z}{u}\right)^ {\frac{n}{2}+s}\
\mathbf{K}_{\frac{n}{2}+s}(zu)
\end{split}
\end{equation}
where $\mathbf{K}_\mu$ is a Bessel function (see for instance
\cite{Gel}), and
\begin{equation}
\begin{split}
\int_0^\infty z^\mu\ \mathbf{K}_{\nu}(zu)\ dz =2^{\mu-1}\
u^{-\mu-1}\ \Gamma\left( \frac{1+\mu+\nu}{2}\right) \Gamma\left(
\frac{1+\mu-\nu}{2}\right),
\end{split}
\end{equation}
(see for example \cite{rus}).

Finally, we thus get the following  expression for
$H_{-n-s+\ell}(x,u)$:
\begin{equation}\label{H's}
\begin{split}
&H_{-n-s+\ell}(x,u)
=\dfrac{-i\ 2^{s-2\ell-2}} {\pi^{\frac{n}{2}+1}\ \ell ! }
e^{-i\frac{\pi}{2}s}\ \sin(\pi s)\\ \\& \times( -i\not
\!\partial_u+\partial_a)\ \left[\ \not \!\!D_x \ ( -i\not
\!\partial_u+\partial_a)\right]^\ell
\sum^{\ell+1}_{k=0}\frac{(-ia)^k}{k!}\\ \\& \times
\left.\Gamma\left( \frac{1+s+k}{2}\right) \Gamma\left(
\frac{s+k+n-1-2\ell}{2}\right)\ u^{-s-n+2\ell+1-k}\ \right|_{a=0}.
\end{split}
\end{equation}
The first four terms $H_{-n-s+\ell}(x,u)$'s, obtained from
(\ref{H's}) after a straightforward but tedious computation  just
involving $\gamma$-matrices's algebra and derivatives, are shown
in Table~I. There, as usual, $F_{\mu\nu}=
\partial_\mu A_\nu -\partial_\nu A_\mu=-i(D_\mu A_\nu -D_\nu A_\mu)$. It is worth
noticing that the first terms of the exponential
\begin{equation}\begin{split}
e^{-i\int_x^y A.dz}  = 1 +
i(u.A)-\frac{(u.D)(u.A)}{2!}-i\frac{(u.D)(u.D)(u.A)}{3!}+\dots
\end{split}\end{equation}
start to appear as an overall  factor  in the sum of the expansion
(\ref{01}) for $K_s(x,y)$.

\begin{table}[!]\label{tabla}
\caption{The first four $H_{-n-s+\ell}(x,u)$'s. } \centerline{
\begin{tabular}{|l|}\hline \\
{$H_{-n-s}(x,u)=\dfrac{2^{s-1}}{\pi^{\frac{n}{2}+1}}\, \, \,
e^{-i\frac{\pi}{2}s}\, \sin(\pi s)$\hfill }\\ \\ $\times \left[\
\Gamma\left(\frac{1+s}{2}\right)
\Gamma\left(\frac{n+s+1}{2}\right) u^{-n-s-1} \not \!u \ -\
\Gamma\left(\frac{2+s}{2}\right) \Gamma\left(\frac{n+s}{2}\right)
u^{-n-s}\ \right]$
\\ \\ \hline  \\
$ H_{-n-s+1}(x,u)=\dfrac{2^{s-1}}{\pi^{\frac{n}{2}+1}}\, \, \,
e^{-i\frac{\pi}{2}s}\, \sin(\pi s)$\\ \\ $\times\left[\
\Gamma\left(\frac{1+s}{2}\right)
\Gamma\left(\frac{n+s+1}{2}\right) u^{-n-s-1}  \not \!u\right.
\left.-\ \Gamma\left(\frac{2+s}{2}\right)
\Gamma\left(\frac{n+s}{2}\right) u^{-n-s}\ \right] i(u.A) $\\ \\
\hline  \\
$H_{-n-s+2}(x,u)=\dfrac{2^{s-1}}{\pi^{\frac{n}{2}+1}}\,
\, \, e^{-i\frac{\pi}{2}s}\, \sin(\pi s)\,\, $\\ \\
$\times\biggl\{\left[\ \Gamma\left(\frac{1+s}{2}\right)
\Gamma\left(\frac{n+s+1}{2}\right) u^{-n-s-1}  \not \!u\right.
\left.-\ \Gamma\left(\frac{2+s}{2}\right)
\Gamma\left(\frac{n+s}{2}\right) u^{-n-s}\ \right]
\left(-\frac{(u.D)(u.A)}{2!}\right)\biggr. $\\ \\$
\biggl.+\frac{i}{8}\left[\ \Gamma\left(\frac{1+s}{2}\right)
\Gamma\left(\frac{n+s-1}{2}\right) u^{-n-s+1}\ u_\rho \gamma_\mu
\gamma_\rho \gamma_\nu +\Gamma\left(\frac{2+s}{2}\right)
\Gamma\left(\frac{n+s-2}{2}\right) u^{-n-s+2}\ \gamma_\mu
\gamma_\nu\right]F_{\mu\nu}\biggr\} $\\ \\ \hline \\
$H_{-n-s+3}(x,u)=\dfrac{2^{s-1}}{\pi^{\frac{n}{2}+1}}\, \, \,
e^{-i\frac{\pi}{2}s}\, \sin(\pi s)\,\, $\\ \\$
\times\biggl\{\left[\ \Gamma\left(\frac{1+s}{2}\right)
\Gamma\left(\frac{n+s+1}{2}\right) u^{-n-s-1}  \not \!u\right.
\left.-\ \Gamma\left(\frac{2+s}{2}\right)
\Gamma\left(\frac{n+s}{2}\right) u^{-n-s}\ \right]
\left(-i\frac{(u.D)(u.D)(u.A)}{3!}\right)\biggr. $\\ \\$
\left.+\frac{i}{8}\left[\ \Gamma\left(\frac{1+s}{2}\right)
\Gamma\left(\frac{n+s-1}{2}\right) u^{-n-s+1}\ u_\rho \gamma_\mu
\gamma_\rho \gamma_\nu +\Gamma\left(\frac{2+s}{2}\right)
\Gamma\left(\frac{n+s-2}{2}\right) u^{-n-s+2}\ \gamma_\mu
\gamma_\nu\right]F_{\mu\nu}\, \, i(u.A) \right.
 $\\ \\$ + \frac{1}{24} \left[\ \Gamma\left(\frac{1+s}{2}\right)
\Gamma\left(\frac{n+s-1}{2}\right)
u^{-n-s+1}\left(-\frac{3}{2}u_\rho u_\sigma \gamma_\mu
\gamma_\rho \gamma_\nu \Dc_\sigma F_{\mu\nu}-u_\mu u_\rho
\gamma_\rho
 \Dc_\nu F_{\mu\nu}+u_\mu u_\nu
\gamma_\rho \Dc_\nu F_{\mu\rho} \right)\right.$\\ \\$+ \left.
\Gamma\left(\frac{2+s}{2}\right)
\Gamma\left(\frac{n+s-2}{2}\right)u^{-n-s+2}\left(-\frac{3}{2}u_\mu
\gamma_\nu \gamma_\rho \Dc_\mu F_{\nu\rho}+u_\mu \Dc_\nu
F_{\mu\nu} \right)\right. $\\ \\$ \bigl.
\left.+\Gamma\left(\frac{1+s}{2}\right)
\Gamma\left(\frac{n+s-3}{2}\right)u^{-n-s+3}\ \gamma_\nu \Dc_\mu
F_{\mu\nu}\ \right] \biggr\}$\\ \\  \hline
\end{tabular}}
\end{table}

Now, we shall compute the sum in expression (\ref{chivo}) in
order to see whether both methods coincide or not. Taking into
account that $G_{-n+1+\ell}\ (x,u)= \underset
{s\rightarrow-1}{\lim}\ H_{-n-s+\ell}\ (x,u)$ for $\ell<N$ and
$G_{-n+1+N}\ (x,u)= \underset {s=-1}{\fp}\ H_{-n-s+N}\ (x,u)$,
from Table I we get the following relations.

For $n=2$, we have
\begin{equation}\begin{split}
\sum_{\ell=0}^{1} G_{-2+1+\ell}\ (x,u)\  e^{i\int_x^y A.dz} = -
\frac{i}{2\pi} \frac{\not \!u}{u^2}(1+o(u^2)),
\end{split}\end{equation}
so it is clear that (\ref{chivo}) holds in this case.

For $n=3$, we get
\begin{equation}\begin{split}
\sum_{\ell=0}^{2} G_{-3+1+\ell}\ (x,u)\  e^{i\int_x^y A.dz}  = -
\frac{i}{4\pi} \frac{\not \!u}{u^3}(1+o(u^3))\\ +\frac{1}{16
\pi}\left[\frac{u_\rho}{u} \gamma_\mu \gamma_\rho\gamma_\nu +
\gamma_\mu \gamma_\nu\right]F_{\mu\nu},
\end{split}\end{equation}
and so
\begin{equation}
\Sl \tr \left( \gamma_\mu \sum_{\ell=0}^{2} G_{-3+1+\ell}\ (x,u)\
e^{i\int_x^y A.dz} \right)=\frac{1}{16 \pi} \tr[\gamma_\mu
 \gamma_\rho \gamma_\nu ]F_{\rho\nu},
\end{equation}
which vanishes or not depending on the $\gamma$'s representation
(it does not vanish if the $2\times2$ Pauli matrices are chosen).

Finally, we consider $n=4$. In this case, a pole is present in
$H_{-4-s+3}(x,u)$ at $s=-1$. After computing the finite part in
order to get $G_{-4+1+3}(x,u)$ we have
\begin{equation}\begin{split}
&\sum_{\ell=0}^{3} G_{-4+1+\ell}\ (x,u)\  e^{i\int_x^y A.dz} = -
\frac{i}{2\pi^2} \frac{\not \!u}{u^4}(1+o(u^4))\\&+ \frac{1}{16
\pi^2}\frac{u_\rho}{u^2} \gamma_\mu \gamma_\rho\gamma_\nu
F_{\mu\nu}(1+o(u^2))\\& - \frac{i}{48\pi^2} \frac{u_\rho
u_\sigma}{u^2} ( -\frac{3}{2} \gamma_\mu \gamma_\rho\gamma_\nu
\partial_\sigma F_{\mu\nu} -\gamma_\rho \partial_\mu F_{\sigma\mu}
+\gamma_\mu\partial_\rho F_{\sigma\mu})\\
&- \frac{i}{24\pi^2}  ( \ln 2 - \ln u - \frac{i\pi}{2} +
\Gamma'(1)) \gamma_\nu\partial_\mu F_{\mu\nu},
\end{split}\end{equation}
which, in general, clearly yields  a nonzero result for expression
(\ref{chivo}).

So, we see that although  Schwinger and complex powers methods
are both gauge invariant, they only coincide  for the
two-dimensional case. In $3$ dimensions the coincidence depends
on the representation chosen for the $\gamma$-matrices's, while
for $n=4$ they in general disagree.

\end{document}